\documentstyle[prl,multicol,aps,epsf]{revtex}

\newcommand{\tJ}{$t$-$J$\ }

\newcommand{\nb}{n_b}
\newcommand{\nf}{n_f}
\newcommand{\nhole}{n_{\rm hole}}
\newcommand{\tautr}{\tau_{\rm tr}}
\newcommand{\tauH}{\tau_{\rm H}}
\newcommand{\tauhole}{\tau_{\rm hole}}
\newcommand{\taur}{\tau_b}
\newcommand{\tc}{T_{\rm c}}
\newcommand{\thetaH}{\theta_{\rm H}}

\begin{document}

\draft 

\title{Transport phenomenology for a holon-spinon fluid}
\author{Derek K. K. Lee and Patrick A. Lee} \address{Department of
  Physics, Massachusetts Institute of Technology, Cambridge, MA,
  02139} 
\date{\today} 
\maketitle

\begin{abstract}
  We propose that the normal-state transport in the cuprate
  superconductors can be understood in terms of a two-fluid model of
  spinons and holons. In our scenario, the resistivity is determined
  by the properties of the holons while magnetotransport involves the
  recombination of holons and spinons to form physical electrons. Our
  model implies that the Hall transport time is a measure of the
  electron lifetime, which is shorter than the longitudinal transport
  time. This agrees with our analysis of the normal-state data.  We
  predict a strong increase in linewidth with increasing temperature
  in photoemission. Our model also suggests that the AC Hall effect is
  controlled by the transport time.
\end{abstract}


\begin{multicols}{2}
\narrowtext

The normal state of the cuprate superconductor exhibits anomalous
transport properties \cite{ongrev}. In this paper, we discuss the
implications of the experimental results for theoretical ideas based
on spin-charge separation in this system. We believe that the
longitudinal transport may be described by a boson-only theory for
the charge degree of freedom \cite{monte}, while transport in a
magnetic field is controlled by spin-charge recombination.

We review here the experimental results which provide severe
constraints on possible theories. We first focus on the case of optimal
doping where the superconducting transition temperature $\tc$ is
highest. The in-plane resistivity is linear in temperature $T$. The
relaxation rate, measured from a Drude-like peak in the optical
conductivity, appears to be universal
\cite{orenstein,gaoLaSr,romeroBi}:
\begin{equation}
  \label{tautr}
  \hbar/\tautr \simeq 2k_{\rm B}T \; .
\end{equation}
The spectral weight under the Drude-like peak (or derived from the
London penetration depth) is proportional to the hole doping $x$ and
can be written as $e^2x/ma^2$, where $a$ is the lattice constant and
$m$ is found to be close to twice the free electron mass. In a
tight-binding model, this mass corresponds to a hopping integral of
1540K. This is close to the antiferromagnetic exchange $J$ but can
also be interpreted as $t/3$. The latter interpretation is consistent
with recent studies of the \tJ model \cite{jaklic}. The Hall
coefficient, on the other hand, is found to be suppressed from the
classical value $1/xnec$ \cite{chien}. It rises as $1/T$ with
decreasing $T$, approaching the classical value only near $\tc$. The
magnetoresistance \cite{harris} is also strongly suppressed from the
expectation that $\Delta\rho/\rho$ scales as $\tautr^2$.

There has been remarkable success in analyzing the transport data
\cite{ongrev,chien,harris} by introducing a new Hall time scale
$\tauH$, as suggested by Anderson \cite{pwa}. In this picture, the
Hall angle $\thetaH=\sigma_{xy}/\sigma_{xx}$ and $\Delta\rho/\rho$ are
given by:
\begin{eqnarray}
  \label{hall}
  \tan\thetaH = \omega_c\tauH,&\quad& 
  \Delta \rho / \rho \simeq (\omega_c\tauH)^2,\nonumber\\
  \hbar/\tauH &\simeq& T^2/W_{\rm H} \; ,
\end{eqnarray} 
where $\omega_c=eB/mc$ is the cyclotron frequency and $W_{\rm H}$ is
a temperature scale which we discuss below. In the presence of
non-magnetic Zn impurities \cite{chienZn}, $1/\tauH$ extrapolates to a
finite value at zero temperature but its temperature dependence is
unaffected. This residual value is proportional to the impurity concentration
and is comparable to its longitudinal counterpart.  This suggests that
$1/\tauH$ is more than a fitting parameter and represents a physical
process.

Consider now the temperature scale $W_{\rm H}$. For 90K YBCO, $\thetaH^2=
DB^2/T^4$ where $D=$1630 K$^4$T$^{-2}$ \cite{harris}.  From
(\ref{hall}), we see that the Hall angle measures $\tauH/m$.  Using
the mass extracted from the optical spectral weight, we obtain $W_{\rm H} =
(\hbar/2 e a^2)(k_{\rm B}D^{1/2}/J)\simeq 65{\rm K}$ for 90K YBCO. The
Hall time $\tauH$ is therefore {\em shorter} than the longitudinal
time $\tautr$ above 130K, {\em i.e.\/} in the whole of the normal
state except for the region close to $\tc$. Under the assumption of a
single mass, the Hall coefficient is $R_{\rm H}\simeq (1/xnec)
\tauH/\tautr$ so that its reduction from the classical value is direct
evidence that $\tauH<\tautr$.  We note that this analysis is different
from the original analysis of Refs.~\cite{chien,pwa}, which assumes
that $\tauH$ is controlled by the decay of a long-lived quasiparticle
($W_{\rm H}\simeq J$) so that it is longer than the transport time $\tautr$.
As recognized by these authors, this leads one to deduce a carrier
mass 20 times larger than the one used above.

The dynamical time for the decay of Hall currents can be obtained in
the AC Hall effect. For 90K YBCO, Ref.\ \cite{kaplan} gives a ratio
between 2 and 4 for $\tauH^{\rm ac}/\tautr$ at 95K while Ref.\ 
\cite{spielman} gives a lower limit of order unity. The proximity to
$\tc$ makes one worry whether this is characteristic of the normal
state. Measurements at higher temperatures would be desirable.

The discussion so far has been concerned with the optimally-doped
cuprates. In the underdoped regime, the physics is complicated by the
existence of a spin gap. This pseudogap causes a reduction in the
resistivity and $1/\tautr$ by roughly a factor of two, but $\cot\thetaH$
remains quadratic in $T$ with a small increase in $W_{\rm H}$
\cite{uchida}.  In overdoped samples, both the Hall angle and the
resistivity have quadratic temperature dependences \cite{mackenzie}.
Although a scattering rate of $T^2/W$ is in accordance with
Fermi-liquid theory, we observe that the temperature scale $W$ is much
smaller than the bandwidth $J$. To estimate $W$, we note that the
resistivities of overdoped Tl$_2$Ba$_2$CuO$_{6+\delta}$ and optimally
doped YBCO cross at room temperature \cite{mackenzie}. Optical data
\cite{timusk} indicate that the spectral weight of the overdoped Tl
compound is similar to that of the optimally doped materials, so that the
resistivity ratio between the two compounds is a good indication of
the lifetime ratio.  We therefore estimate that $W\simeq$ 150K in the
overdoped compound. Similarly, $\cot\thetaH$ for Tl is 60\% smaller
than for YBCO. This yields an estimate of $W_{\rm H}\simeq$ 110K. Both
$W$ and $W_{\rm H}$ are much smaller than the bandwidth,
indicating that, even in overdoped materials, the scattering mechanism
is not the conventional screened Coulomb interaction between
electrons.

Anderson has emphasized that the appearance of $\tauH$ may be a
signature of spin-charge separation. There have been attempts to
derive (\ref{hall}) based on the Boltzmann transport of a single
carrier with unusual scattering mechanisms
\cite{stojkovic,coleman,kotliar}. For example, Coleman {\em et
  al.\/}\cite{coleman} introduce a mechanism which does not conserve
particle number. Kotliar {\em et al.\/} \cite{kotliar} use a skew
scattering rate which diverges at low temperatures, and $\tauH$
appears not as a physical rate but as a ratio of two rates so that its
behavior with impurity is difficult to rationalize. In this paper, we
abandon the notion of a single carrier, and explore a
phenomenology based on spin-charge separation.

We review first the picture of spin-charge separation in the \tJ model
which, we believe, describes the low-energy physics of the cuprates.
In the slave-boson treatment \cite{ioffe89,naglee}, the introduction
of a physical hole (of spin $\sigma$ at site $i$) away from
half-filling is represented as the creation of a charged hard-core
boson (holon) and the destruction of a neutral spin-half fermion
(spinon): $c_{i\sigma}=b_i^{\dagger}f_{i\sigma}$, with a
single-occupancy constraint: $b_i^{\dagger}b_i + \sum_\sigma
f_{i\sigma}^{\dagger}f_{i\sigma}= 1$. For a doping of $x$ holes per
site, the holon and spinon densities are $\nb=x$ and $\nf=1-x$
respectively. In the uniform resonating-valence-bond ansatz,
short-range antiferromagnetic correlations are incorporated into the
model by assuming that $\sum_{\sigma}\langle
f_{i\sigma}^{\dagger}f_{j\sigma}\rangle =\xi e^{ia_{ij}}$. At the
mean-field level, there is no net gauge flux ($a_{ij}=0$) so that the
holons have a bandwidth controlled by the hopping integral $t$ of the
original electrons while the spinon bandwidth is controlled by the
antiferromagnetic exchange $J$. In this paper, we will focus on the
cuprates near optimal doping where this slave-boson scheme is believed
to apply. For instance, it gives rise to a large Fermi surface, as
observed in photoemission experiments.

The fluctuations in the gauge field $a_{ij}$ are strong. For
temperatures above the experimental $\tc$, the transverse part of the
fluctuations corresponds to a magnetic field with a root-mean-square
value of the order of a flux quantum per plaquette \cite{monte}. These
fluctuations arise because the single-occupancy constraint requires
that the spinon and holon number currents cancel each other:
\begin{equation}
  \label{constraint}
  {\bf J}_{\rm f} + {\bf J}_{\rm b} = 0\;.
\end{equation}

At sufficiently low temperatures, the bosons become phase-coherent,
leading to well-defined physical electron quasiparticles, {\em i.e.\/}
spinon-holon confinement or the breakdown of spin-charge separation.
We believe that this confinement occurs at $\tc$, consistent with the
fact that the electronic quasiparticles are long-lived in the
superconducting state \cite{spielman}.

Consider now the effect of the gauge field on the transport properties
of the system. Longitudinal transport should be dominated by the
dynamics of the charged holons. The holons are strongly scattered by
the internal magnetic fields which can be regarded as quasistatic
disorder at long wavelengths and low temperatures. We have shown in a
quantum Monte Carlo study \cite{monte} that this gives rise a holon
scattering rate equal to $2k_{\rm B}T$ which should also be the
scattering rate relevant to longitudinal transport. This is consistent
with the relaxation rate (\ref{tautr}) deduced from the optical
conductivity.

The picture that emerges from our study is that, in the normal state,
the boson de Broglie wavelength is much larger than the interparticle
spacing so that the bosons undergo strong exchange and should be
viewed as a quantum liquid rather than single particles. The strong
gauge field forces the boson world lines to retrace each other, and
prevents the development of a superfluid density. This Bose liquid is
insensitive to magnetic fields because retracing paths do not detect
any Aharonov-Bohm phase. The holon fluid therefore has negligible Hall
effect and magnetoconductivity. In the random-phase approximation, the
total Hall coefficient of the spinon-holon fluid is given by the
Ioffe-Larkin rule \cite{ioffe89,naglee}:
\begin{equation}
  \label{hallioffe}
  R_{\rm H} = \frac{\chi_{\rm f}R_{\rm H,b}+
  \chi_{\rm b}R_{\rm H,f}}{\chi_{\rm f}+\chi_{\rm b}}\; ,   
\end{equation}
where $R_{\rm
  H,b}$ and $\chi_{\rm b}$ are the holon Hall coefficient and orbital
susceptibility and $R_{\rm H,f}$ and $\chi_{\rm f}$ are the
corresponding quantities for the spinons. We therefore see that the spinon
contribution to the Hall response is also small, since the orbital
susceptibility of the bosons is suppressed by the gauge fluctuations
for the same reason that their Hall response is suppressed.

It is possible that the self-retracing approximation breaks down due
to gauge-field dynamics or a reduction in gauge amplitude so that the
response to a magnetic field is gradually restored at low
temperatures. In this paper, we explore another possibility. We
suggest that the retracing picture remains valid down to $\tc$ so that
the magnetic response is beyond the scope of a holon-only model.
Instead, we propose that the magnetic response could be understood in
terms of the incipient recombination of the holons and spinons. This
is based on the observation that the physical hole does not experience
any fictitious gauge fields so that its magnetic response should not
be suppressed. In other words, the Aharonov-Bohm phases of the holons
and the spinons due to the internal gauge field now cancel each other, and
the physical hole is not self-retracing. As already mentioned, the
Hall coefficient of the cuprates indeed approaches the classical value
as one approaches $\tc$ where we believe spin-charge confinement to be
complete. We will now develop a simple phenomenogy for normal-state
transport based on these ideas. We do not claim to have the final
answer, because we are forced to make a number of assumptions before
we can arrive at (\ref{hall}).  We can only give some indication of
how one might justify some of the assumptions in terms of the gauge
theory of the \tJ model.

In the normal state above $\tc$, the charge carrier exists in two
states --- either as a holon or as a physical hole:
\begin{equation}
  {\rm holon(b)+antispinon(\overline{f})}\rightleftharpoons{\rm hole(h)}
\end{equation}
The densities of the holons $\nb$, spinons $\nf$ and physical holes
$\nhole$ obey: $\nb+\nhole=x$ and $\nf+\nb=1$ where $x$ is the doping
per site. The carrier exists as a holon for a time interval of the
order of $\taur$ before recombining with an antispinon to form a
physical hole. This physical hole has a lifetime of $\tauhole$ before
decaying back into its spin and charge components. By the principle of
detailed balance, the equilibrium densities obey:
\begin{equation}
\frac{\nb \nf}{\taur} = \frac{\nhole}{\tauhole} \;.
\end{equation}
In a regime of spin-charge separation, spin-charge recombination is
rare and the electron lifetime is short, {\em i.e.}, $\tauhole \ll
\taur$, so that $\nhole \ll \nf,\nb$, and $\nhole\simeq
x\tauhole/\taur$. As one approaches the confinement regime, $\tauhole$
becomes larger than $\taur$, and $\nhole\simeq\nb$.

We will now discuss the implications of this scenario for transport
properties. As mentioned above, the charge carrier responds to
external magnetic fields only as a physical hole. Consider a simple
classical model where the response of the charge carriers to an
external magnetic field is switched on for a duration of $\tauhole$
and switches off for a duration of $\taur \gg \tauhole$, corresponding
to the deconfined and confined states respectively. An electric field
$E$ in the $x$-direction accelerates a particle for a duration of
$\tautr$ before the particle velocity is randomized. Thus, the drift
velocity of the system is $v_x \sim eE\tautr/m$, and $\sigma_{xx} =
ne^2\tautr/m$ where $m$ is the holon mass in the spin-charge-separated
regime. In this time interval, a particle also receives on average
$\tautr/\taur$ impulses of $ev_x B \tauhole/c$ in the $y$-direction due
to the Lorentz force. The transverse drift momentum is $mv_y
\sim (ev_xB/c)(\tautr/\taur)\tauhole=eE\omega_c\tautr^2\tauhole/\taur$.
From this, we deduce a Hall angle of $\theta_H \simeq v_y/v_x =
\omega_c \tauH$ where
\begin{equation}
  \tauH \simeq \frac{\tauhole}{\taur}\tautr \; .
  \label{tauH}
\end{equation}
We therefore see that the Hall effect is reduced from the Fermi-liquid
result ($\tauH=\tautr$) by a fraction of
$\tauhole/\taur\simeq\nhole/\nb$. One can also see in this picture
that the $x$-component of the Lorentz force gives rise to a
negative magnetoconductivity proportional to $(\omega_c \tauH)^2$.

More concisely, we have a model where the drift velocity ${\bf v}$
obeys the following dynamics:
\begin{equation}
  \label{relax}
  m\dot{\bf v} + \frac{m{\bf v}}{\tautr} = 
  e{\bf E} + \frac{e}{c} \eta(t) {\bf v}\times{\bf B}\;. 
\end{equation}
The random function $\eta(t)$ is zero except for spikes of value unity
and duration $\tauhole$. These spikes occur with a time spacing of the
order of $\taur$. Therefore, at time scales greater than $\taur$, the
system sees an effective reduction in the magnetic field by a factor
of $\overline{\eta}{\bf B}$, where $\overline{\eta}=\tauhole/\taur$ is
the time-averaged value of $\eta$.  It should be noted that the simple
model (\ref{relax}) does not involve separate decay rates for the
longitudinal and transverse drift velocities so that the width
$1/\tauH^{\rm ac}$ of the AC Hall angle $\thetaH({\omega})$ is given
by the transport relaxation time $1/\tautr$ rather than $1/\tauH$.
This provides an important test of our hypothesis. As mentioned above,
current experimental data give $\tauH^{\rm ac}$ to be of the same
order of magnitude as $\tautr$ just above the superconducting
transition.  To settle this issue, it would be necessary to measure
the full temperature dependence of the AC Hall relaxation rate.

\begin{figure}
  \begin{center}
    \leavevmode
    \epsfxsize=0.8\columnwidth\epsfbox{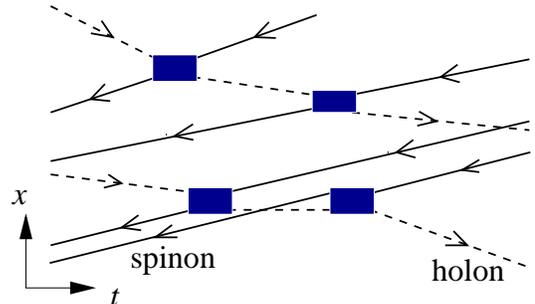}
    \caption{Schematic picture of recombination and decay between spinons
      (solid line) and holons (dashed line) and physical holes
      (box). A hole lives for a time $\tauhole$, shorter
      than the holon lifetime $\taur$ and the spinon lifetime
      $\taur/x$. Only the physical hole experiences an external
      magnetic field.}
    \label{fig:scatter}
  \end{center}
\end{figure}
An important assumption behind this phenemenological model is that,
whereas the average charge current is relaxed by the gauge-field
scattering, the binding of a holon with an antispinon and the
subsequent decay of the physical hole do not provide an additional
mechanism to dissipate momentum from the holon-spinon system. Note
that this argument requires only the total drift current to be
conserved during recombination and decay. One might ask how a holon
(which carries the charge) might have memory of its pre-recombination
velocity when it is re-emitted upon the decay of the physical hole.
From the view of the holon which carries a small momentum, it would
appear that its momentum is strongly affected in each decay and
recombination process. However, from the view of the spinon, which
carries a large momentum of order $\pi/a$, it is reasonable that it is
scattered mainly in the forward direction and that its velocity is
preserved. This is indicated in Fig.\ \ref{fig:scatter}.  We can now
appeal to the current constraint (\ref{constraint}) to argue that,
{\em on average}, the boson current is also conserved. The constraint
is relaxed locally in a spin-charge separated system, but must remain
in force on larger length scales. Another way of saying this is that
Fig.\ \ref{fig:scatter} is misleading in the sense that the holons are
strongly overlapping and exchanging and should not be viewed as
individual particles.

So far, our phenomenological model contains three time scales whereas
there are only two independent time scales to observe. However, as
already pointed out, $1/\tauH$ is a physical scattering rate, rather
than a combination of time scales as shown in (\ref{tauH}). This
forces us to conjecture further that $\taur$ is of the order of
$\tautr$. A consequence of this conjecture is that:
\begin{equation}
  \tauH \sim \tauhole \; .
\end{equation}
In other words, the Hall transport time is a measure of the lifetime
of the physical hole. This provides another important test of our
model. The hole lifetime can be deduced independently from
angle-resolved photoemission linewidths which we predict to grow as
$T^2/W_{\rm H}$. The small size of $W_{\rm H}$ leads to a severe broadening at
room temperature which should be amenable to experiments.

The assumption that $\taur\sim\tautr$ is the weakest point of our
argument. The only justification we can offer is that, due to the
mismatch of the kinematics of the spinon and the holon, the
recombination process is perhaps controlled by the same momentum
relaxation process which contributes to $\tautr$. We also need to
argue that, in the presence of impurities, $1/\tauhole$ becomes
$1/\tauhole + 1/\tau_0$ where $1/\tau_0$ is a residual value due to
impurity scattering. This is not unreasonable in that a hole may well
disintegrate rapidly on encountering an impurity. Finally, we can
offer no explanation of why $\tauhole$ should scale as $1/T^2$.  More
importantly, we do not understand the origin of the temperature scale
$W_{\rm H}$ which is small and not very sensitive to doping. It is
particularly puzzling that the spin gap in the underdoped
cuprates has a much smaller effect on $\tauH$ than on $\tautr$.

We will now comment briefly on the overdoped regime. We have already
pointed out that the resistivity, albeit quadratic in temperature, is
much too high compared to Fermi-liquid theory. We propose that
$\tauhole$ increases with doping so that, in the overdoped regime,
there is a region of temperature above $\tc$ where the physical hole
is more stable and $\tauhole>\taur$. In this case, although the
current is dissipated in the holon state, the rate-determining step
for current relaxation is the conversion of the physical hole into a
holon and an antispinon. We therefore argue that the resistivity is
controlled in this regime by $1/\tauhole\sim 1/\tauH$ and not the
holon scattering rate. Upon further doping, the system eventually
crosses over to Fermi-liquid theory when $\tauhole$ becomes comparable
to the Fermi-liquid scattering time.

In summary, we have put forward a hypothesis for understanding
transport in the cuprates, based on the idea of spin-charge
reconfinement. This model explains naturally the suppression of the
Hall response compared to the classical Drude theory. It also links
the Hall transport time $\tauH$ in DC measurements to the lifetime of
a physical hole, while the dynamical time extracted from the AC Hall
effect is expected to be the longitudinal scattering time $\tautr$. We
hope that our model, while incomplete, will stimulate further
experimental work and serve as the basis for further discussion. This
work is supported primarily by the NSF MRSEC program (DMR-9400334).

\end{multicols}

\begin{references}
  
\bibitem{ongrev} For a review, see N.P.~Ong, Y.F.~Yan and J.M.~Harris,
  in {\em Proceedings of the CCAST Symposium on High-$T_{\rm c}$
    Superconductivity and the C$_{60}$ Family, Beijing 1994\/} (Gordon
  and Breach, New York, 1995).
  
\bibitem{monte} D.K.K.~Lee, D.H.~Kim and P.A.~Lee, Phys. Rev. Lett.
  {\bf 76}, 4801 (1996); D.H.~Kim, D.K.K.~Lee and P.A.~Lee, to be
  published in Phys. Rev.  B (1996).


\bibitem{orenstein} J.~Orenstein {\em et al.\/}, Phys. Rev. B {\bf
    42}, 6342 (1990).

\bibitem{gaoLaSr} F.~Gao {\em et al.}, Phys.~Rev.~B {\bf 47}, 1036
  (1993).

\bibitem{romeroBi} D.B.~Romero {\em et al.\/}, Phys.~Rev.~Lett. {\bf
    68}, 1590 (1992).

\bibitem{jaklic} J.~Jakli\v{c} and P.~Prelov\v{s}ek, Phys. Rev. B {\bf 52},
6903 (1995).

\bibitem{chien} T.R.~Chien, D.A.~Brawner, Z.Z.~Wang and N.P.~Ong,
  Phys. Rev. B {\bf 43} 6242 (1991).

\bibitem{harris} J.M. Harris {\em et al.}, Phys. Rev. Lett. {\bf 75},
  1391 (1995).

\bibitem{pwa} P.W.~Anderson, Phys. Rev. Lett. {\bf 67}, 2092 (1991).

\bibitem{chienZn} T.R.~Chien, Z.Z.~Wang and N.P.~Ong, Phys. Rev.
  Lett. {\bf 67}, 2088 (1991).

\bibitem{kaplan} S.G.~Kaplan {\em et al.\/}, Phys. Rev. Lett. {\bf
    76}, 696 (1996).
  
\bibitem{spielman} S.~Spielman {\em et al.}, Phys. Rev. Lett.
  {\bf 73}, 1537 (1994); B.~Parks and J.~Orenstein, private communication.
  
\bibitem{uchida} S.~Uchida, K.~Takenaka and K.~Tamasaku, {\em Proceedings
    of the 10th Anniversary HTS Workshop, Houston\/} (1996).

\bibitem{mackenzie} A.P.~Mackenzie, S.R.~Julian, D.C.~Sinclair and
  C.T.~Lin, Phys. Rev. B {\bf 53}, 5848 (1996).

\hyphenation{koles-ni-kov}
\bibitem{timusk} A.V.~Puchkov, P.~Fournier, T.~Timusk and
  N.N.~Kolesnikov, Phys. Rev. Lett. {\bf 77} 1853 (1996).

\bibitem{stojkovic} B.P.~Stojkovi\'c and D.~Pines, Phys. Rev. Lett. {\bf
    76}, 811 (1996).

\bibitem{coleman} P.~Coleman, A.J.~Schofield and A.M.~Tsvelik, Phys.
  Rev. Lett. {\bf 76}, 1324 (1996).
  
\bibitem{kotliar} G.~Kotliar, A.~Sengupta and C.M.~Varma, Phys. Rev.
  B {\bf 53}, 3573 (1996).
  
\bibitem{ioffe89} L.B.~Ioffe and A.I.~Larkin, Phys. Rev. B {\bf 39},
  8988 (1989).
  
\bibitem{naglee} N.~Nagaosa and P.A.~Lee, Phys. Rev. Lett. {\bf 64},
  2450 (1990); Phys. Rev. B {\bf 46}, 5621 (1992).
  

\end{references}
\end{document}